\documentclass[preprint,showpacs,preprintnumbers,amsmath,amssymb,endfloats]{revtex4-2}
\usepackage{graphicx}
\usepackage{dcolumn}
\usepackage{bm}
\usepackage{amssymb}
\usepackage{amsmath}
\usepackage{latexsym}
\usepackage{longtable}
\usepackage{color}

\begin{document}

\title{New type of self-oscillating systems}

\author{V.V. Sargsyan$^{1}$, A.A. Hovhannisyan$^{1,2,3}$, G.G. Adamian$^{1}$,  N.V. Antonenko$^{1,4}$, and D. Lacroix$^{5}$ }
\affiliation{
$^{1}$Joint Institute for Nuclear Research, 141980 Dubna, Russia\\
$^{2}$Institute of Applied Problems of Physics,  0014 Yerevan, Armenia\\
$^{3}$Quantum Computing Laboratory,   1142 Norakert, Armenia \\
$^{4}$Tomsk Polytechnic University, 634050 Tomsk, Russia\\
$^{5}$Universit{\'e}
Paris-Saclay, CNRS/IN2P3, IJCLab, 91405 F-91406 Orsay Cedex, France
}
\affiliation{}
\date{\today}

\begin{abstract}
The time evolution of occupation number is studied for a bosonic oscillator (with one and two degrees of freedom)
linearly fully coupled  to fermionic and bosonic heat baths. The absence of equilibrium in this oscillator  is  discussed as a tool to create
a dynamical non-stationary memory storage. The connection between such a system and the well-known nonlinear self-oscillating systems is demonstrated.
%
%
%
%
%
%
%
%
\end{abstract}
\maketitle

\section{Introduction}
%

In Refs. \cite{PhysicaA2019,PRE2020,PRE2020n},
it has been illustrated that a fermionic oscillator (the two-level system) or bosonic oscillator
linearly fully coupled  to several baths of different statistics (fermionic and bosonic)
might never reach a stationary asymptotic limit and the occupation number  in this system oscillates at large times.
%
The period of asymptotic oscillations depends on the frequency of oscillator but not on the environment.
Such a system with non-stationary asymptotics can be used as a dynamical (non-stationary) memory system because
the information about some properties of the system (population of excitation states and frequency) is preserved at large times
and is stable under   external and environmental conditions \cite{PRE2020n}.
%
%
For quantum computers, the imposition of an unambiguous population of excited states is also important.
In this case, one should ensure a sufficient degree of
meta-stability of the excited states of the quantum register.
These states  must have  sufficiently large lifetime that determines their
relaxation to the ground state due to dissipative processes.


%
%
In the present paper, we analyze the relationship between the non-stationary system considered and self-oscillating systems,
and study the time evolution of occupation numbers of two linearly
coupled bosonic oscillators embedded in the fermionic and bosonic heat baths.
The systems under consideration  are linearly fully coupled  to the heat baths of different statistics in the case of Ohmic dissipation with Lorenzian cutoffs.
The linear full coupling case contains the resonant [the rotating wave approximation] and non-resonant terms \cite{M1,Armen,Stefanescu}.
The  environmental effects  on a quantum
system could keep this system in certain state or provide it some specific properties \cite{M1,Armen,Stefanescu}.



\section{Bosonic oscillator linearly fully coupled with  fermionic and bosonic baths}
%
%
%
\begin{figure}
\begin{center}
\includegraphics[scale=0.85,angle=0]{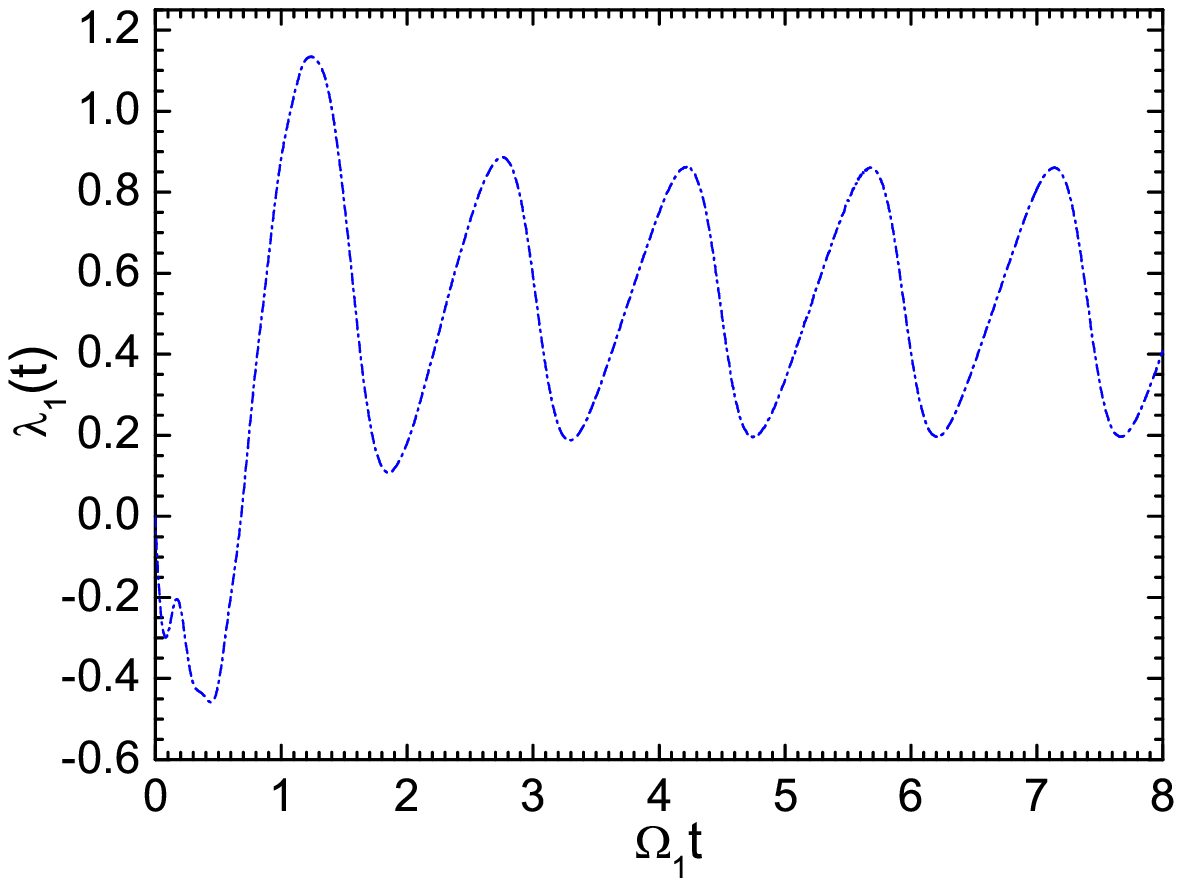}
\includegraphics[scale=0.85,angle=0]{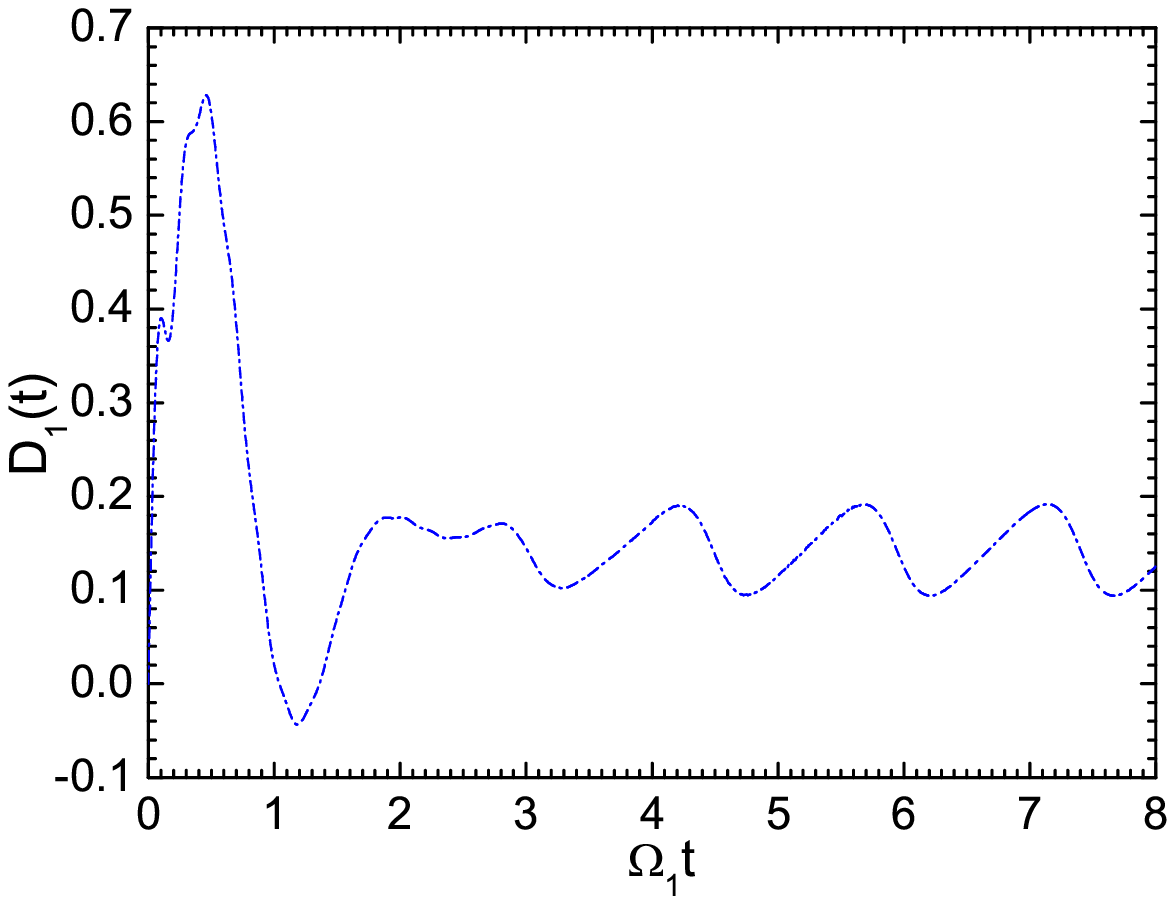}
\end{center}
\vspace{1cm}
\caption{
The calculated dependencies of the friction  $\lambda_1(t)$    and  diffusion    $D_1(t)$   coefficients on time $t$ for the
 bosonic oscillator
in the case of Ohmic dissipation with Lorenzian cutoffs.
The baths  have the same
level densities  with cut-off parameters (the inverse memory times)  $\gamma_{1}/\Omega_1=10$, $\gamma_{2}/\Omega_1=15$.
Here, the coupling strengths between the oscillator and baths are $\alpha_{1}=0.1$ and $\alpha_{2}=0.05$,
temperatures $kT_1/(\hbar\Omega_1)=1$   and $kT_2/(\hbar\Omega_1)=0.1$, and
the renormalized frequency of the bosonic oscillator is $\Omega_1$ \cite{PRE2020n}.
}
\label{fig:l}
\end{figure}
\begin{figure}[h]
\includegraphics[scale=0.85]{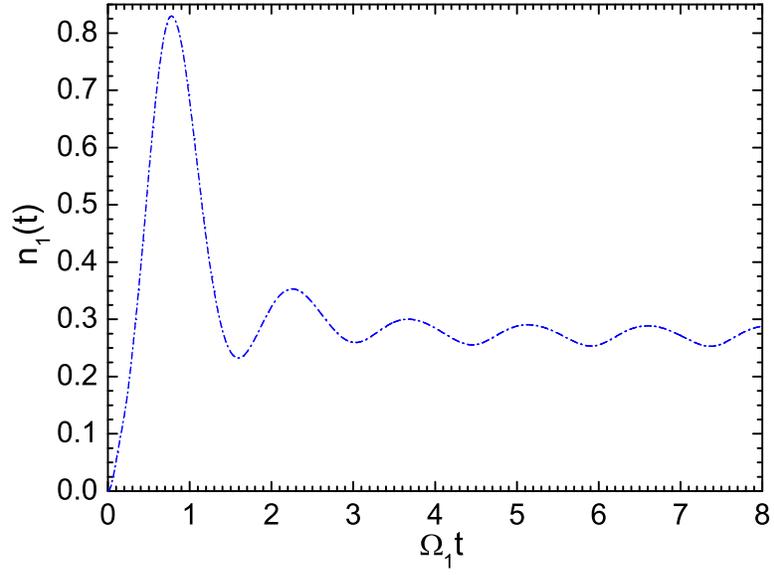}
\caption{
The calculated dependence of the average occupation number $n_1(t)$ on time $t$ for the bosonic oscillator.
The plot   corresponds to   initially unoccupied, $n_1(0)=0$,  oscillator state.
}
\label{fig:2}
\end{figure}
\begin{figure}
\begin{center}
\includegraphics[scale=0.85,angle=0]{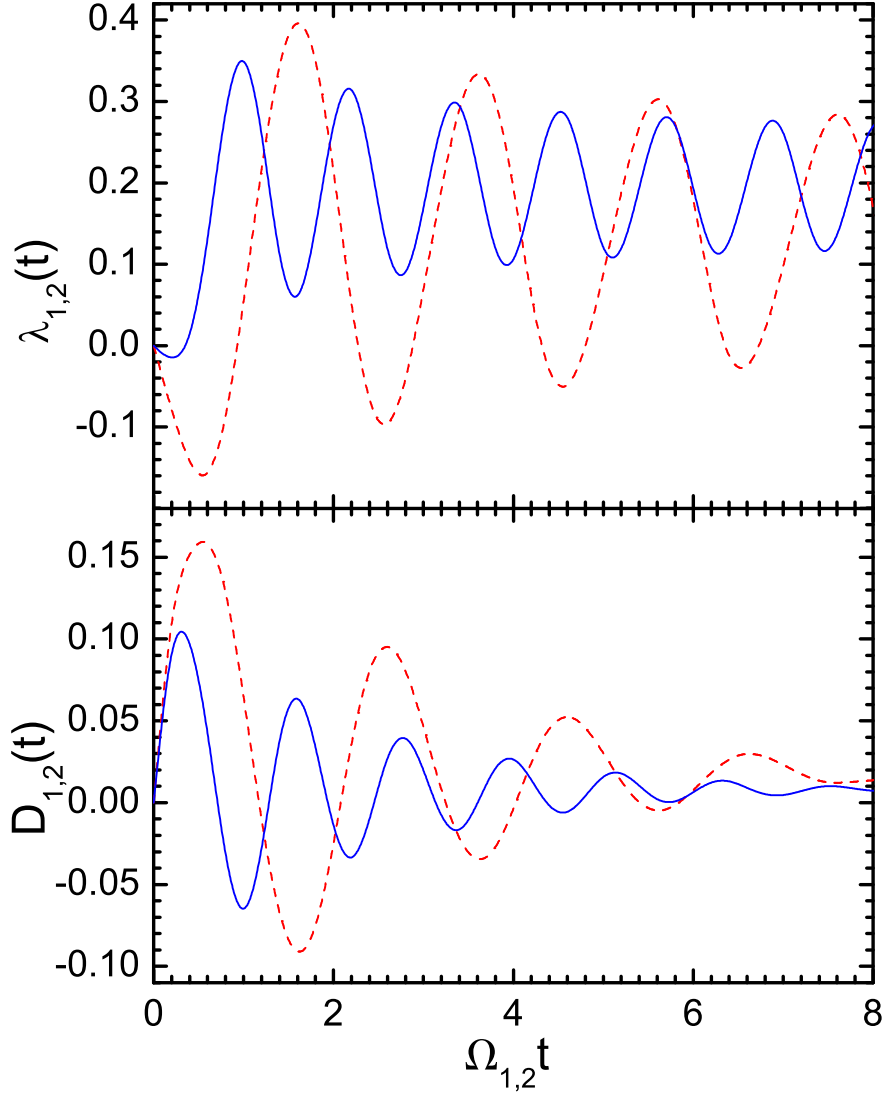}
\end{center}
\vspace{1cm}
\caption{
The calculated dependencies of the friction $\lambda_{1,2}(t)$    and  diffusion    $D_{1,2}(t)$   coefficients on time $t$ for the
two bosonic oscillators (dashed (first oscillator)  and solid (second oscillator)   lines)
in the case of Ohmic dissipation with Lorenzian cutoffs.
The baths  have the same
level densities  with cut-off parameters $\gamma_1/\Omega_1=\gamma_2/\Omega_1=12$ and $\gamma_1/\Omega_2=\gamma_2/\Omega_2=6$.
Here, the coupling strengths  between both oscillators and baths are the same, $\alpha_1=\alpha_2=0.03$,
temperatures $kT_{1}/(\hbar \Omega_{1})=kT_{2}/(\hbar \Omega_{1})=0.5$  and $kT_{1}/(\hbar \Omega_{2})=kT_{2}/(\hbar \Omega_{2})=0.25$, and
the renormalized frequencies of the bosonic oscillators are $\Omega_{1,2}$.
}
\label{fig:3}
\end{figure}
\begin{figure}[h]
\includegraphics[scale=0.75]{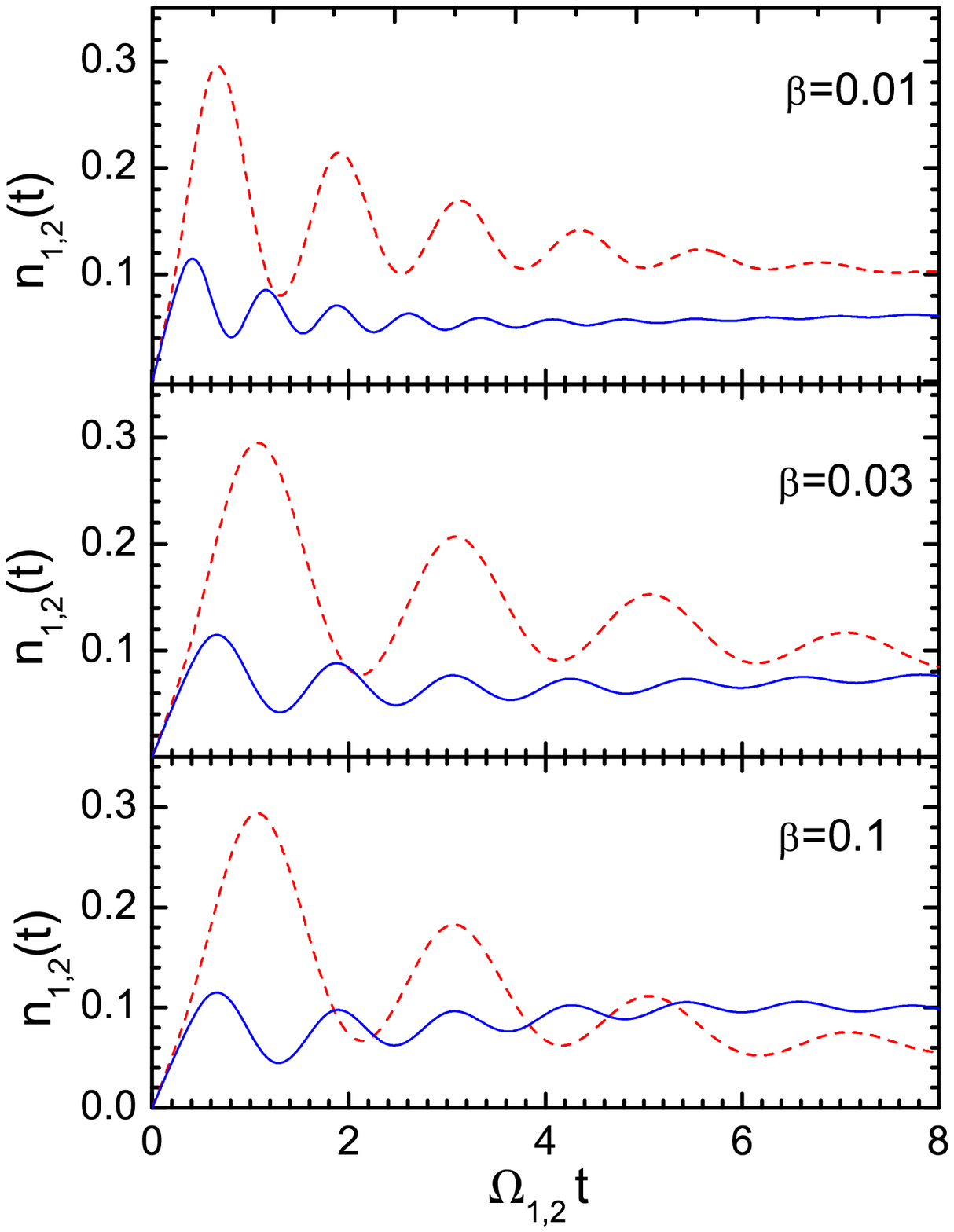}
\caption{
For two bosonic oscillators (dashed (first oscillator)  and solid (second oscillator)   lines),
the calculated dependencies of the average occupation numbers $n_{1,2}(t)$ on time $t$ at different indicated coupling strengths $\beta$ between the oscillators.
The plots  correspond to   initially unoccupied, $n_{1,2}(0)=0$,  oscillator states and $dn_{1,2}(0)/dt=0$.
}
\label{fig:4}
\end{figure}
%
%
%
%
As shown in Ref.  \cite{PRE2020n},
for  bosonic  oscillator (with the renormalized frequency $\Omega_1$)
linearly fully coupled to the fermionic and bosonic
heat baths,  the time evolution of occupation number $n_{1}(t)$ is ruled
by the master-equation
\begin{eqnarray}
\frac{dn_{1}(t)}{dt}=-2\lambda_1(t)n_{1}(t)+2D_1(t),
\label{eq:namaster2}
\end{eqnarray}
where the values  $\lambda_1(t)$ and $D_1(t)$  are the time-dependent friction and diffusion coefficients, respectively.
Explicit expressions for these coefficients are given in Ref. \cite{PRE2020n}.
The master-equation (\ref{eq:namaster2}) can be also rewritten in the following way
\begin{eqnarray}
\frac{d^2n_{1}(t)}{dt^2}+2\lambda_1(t)\frac{dn_{1}(t)}{dt}+2\frac{d\lambda_1(t)}{dt}n_{1}(t)=2\frac{dD_1(t)}{dt}.
\label{eq:namaster2}
\end{eqnarray}

For the  bosonic  oscillator linear fully coupled with the  fermionic and bosonic heat baths,
the time-dependent   friction and diffusion coefficients   are shown in Fig. 1 in the case of Ohmic dissipation with Lorenzian cutoffs.
The diffusion and friction  coefficients are equal to zero at initial time. After quite a short transient time, $\Omega_1 t\le 0.5$,
the friction and diffusion coefficients   oscillate  with the same period of oscillations.
As a result, the  occupation number $n_1(t)$ oscillates around  certain average value at large times, so it has no asymptotic limit (Fig. 2).
The period of these oscillations is determined by the properties of the system but not by the initial condition.
The stable mode of  oscillations is determined by the energy balance, that is,
the equality of the energy lost by the oscillator to the heat bathes and the energy supplied from the heat baths to the oscillator.
As a result, the population of the excited states of system decreases and then increases on the same level independent of the environment.
At $\Omega_1t_1\ge 3$, the period of oscillations of $n_{1}(t)$ at large $t$ is close to $2\pi/\Omega_1$ and, accordingly, carries
information about the system \cite{PRE2020n}.
This gives us a new opportunity to control these oscillations by changing the oscillator frequency \cite{PRE2020n}.







The characteristics of our bosonic oscillator are very similar to those of self-oscillating systems (for example, the well-known Van der Pol oscillator) \cite{Oni}.
The stable mode of self-oscillation is also determined by the energy balance, that is,
the equality of the energy lost by the system and the energy supplied from the external source to the oscillating system. However, our bosonic oscillator
takes energy from the heat baths and gives energy back to them. Our system is more resistant to the noise,
because it directly takes into account the effects of the environment.
Thus, one can say that the bosonic oscillator considered is a new kind of the self-oscillating system.

Note that in general the self-oscillating systems are nonlinear.
These systems are of extremely great interest from the point of view of both fundamental natural science and numerous applications.
There is no single approach to nonlinear methods,
but a number of effective methods have been developed for a wide range of applications \cite{Oni}.
However, their study faces significant difficulties in comparison with linear systems  like  our   system.

\section{Two coupled bosonic oscillators linearly fully coupled with  fermionic and bosonic baths}

In the case of two linearly coupled bosonic oscillators (i.e. with coupling interaction proportional to $i[a^\dagger_1 a_2 -  a_1 a^\dagger_2]$
where $a^\dagger_1$ ($a_1$) and $a^\dagger_2$  ($a_2$) are creations (annihilations) operators of the system $1$ and $2$, respectively \cite{M1})
embedded in the fermionic and
bosonic heat baths, the evolutions of the occupation numbers $n_{1,2}(t)$ are determined by  the solution of the system of two coupled master-equations,
\begin{eqnarray}
\left\{
\begin{array}{l}
\displaystyle
\frac{d^2n_{1}(t)}{dt^2}+2\lambda_1(t)\frac{dn_{1}(t)}{dt}+2\frac{d\lambda_1(t)}{dt}n_{1}(t)+\beta[n_{1}(t)-n_{2}(t)]=2\frac{dD_1(t)}{dt}\\
\\
\displaystyle
\frac{d^2n_{2}(t)}{dt^2}+2\lambda_2(t)\frac{dn_{2}(t)}{dt}+2\frac{d\lambda_2(t)}{dt}n_{2}(t)+\beta[n_{2}(t)-n_{1}(t)]=2\frac{dD_2(t)}{dt}
\end{array}
\right. ,
\label{eq:namaster2-2}
\end{eqnarray}
where  $\beta$  is the coupling strength between two oscillators with the frequencies $\Omega_{1,2}$ and the values of $\lambda_{1,2}(t)$ and $D_{1,2}(t)$
are the time-dependent friction and diffusion coefficients for the first and second systems, respectively. Note that, Eqs. (\ref{eq:namaster2-2})
are obtained assuming that the two systems are coupled to heat-baths that are independent from each others.
In Fig. 3, the values of  $\lambda_{1,2}(t)$ and $D_{1,2}(t)$ oscillate out of phase in the case of Ohmic dissipation with Lorenzian cutoffs.
When interacting, the two bosonic oscillators influence each other, since the connection between them is carried out in both directions.

As seen in Fig. 4,  after   short transient time,  the  occupation numbers $n_{1,2}(t)$ oscillate and they have no asymptotic limits.
As in the case of single bosonic oscillator, the constant periods of these oscillations is determined by the corresponding frequencies
of oscillators but not by their initial conditions and baths.
The frequencies of these oscillations are close to the eigenfrequencies of these oscillators.
The influence of the characteristics of thermal reservoirs of different statistics on the
periods of oscillations of $n_{1,2}(t)$ [$2\pi/\Omega_{1,2}$] is almost negligible. This makes the self-oscillating systems quite robust and
one can control the asymptotic oscillations of both bosonic oscillators by changing their frequencies.
In contrast to the case of a single bosonic oscillator, the oscillation amplitudes change with time $t$
due to the coupling between the oscillators. This change in the amplitude increases with the growth
of the coupling strength $\beta$ between the oscillators (Fig. 4).

%
%
%

The system of two coupled bosonic oscillators is similar to two coupled self-oscillating systems, which have the property of synchronization.
The main feature of synchronization is that two (or more) characteristic time scales of interacting systems, which were independent in the absence of coupling,
are rationally connected  \cite{Oni}. This effect  is also generalized for   an   finite number of coupled bosonic oscillators.
Synchronization for such a system can be used to increase the stability of quantum computers with a dynamical (non-stationary) memory.


\section{Conclusions}
It is shown here that a bosonic oscillator embedded in both fermionic and bosonic heat baths is a
new type of self-oscillating system, where the energy is supplied from the heat baths to the oscillating system.
The fermionic  oscillator linearly fully coupled with
the fermionic and  bosonic heat baths would have the same properties \cite{PRE2020n}.
In the case of two linearly coupled bosonic oscillators embedded in fermionic and bosonic heat baths,
the absence of equilibrium asymptotic  of occupation numbers was predicted.
At large times, the periods of oscillations of occupation numbers mainly depend  on the  frequencies of corresponding oscillators  and, accordingly, carry
information about both oscillators. This is analogous to the synchronization effect of the self-oscillating system.
The similar behavior of the system of two coupled bosonic oscillators can be generalized to a system with an
arbitrary finite number of coupled bosonic oscillators.
Then, such a system is an example of non-stationary (dynamical) memory storage. Each frequency
corresponds to  certain state and can keep the control of these states.
The absence of equilibrium asymptotic  of occupation numbers is also expected for other non-stationary systems
in which  the   asymptotic  friction and diffusion
coefficients periodically oscillate with time and their ratio is not constant.

\section*{Acknowledgments}
%
%
G.G.A. and N.V.A. were supported by  Ministry of Science and Higher Education
of the Russian Federation   (Moscow, Contract No. 075-10-2020-117).
V.V.S. acknowledges the Alexander von Humboldt-Stiftung (Bonn).
D.L. thanks the CNRS for financial support through the 80Prime program and QC2I project.
This work was partly supported by the IN2P3(France)-JINR(Dubna) Cooperation
Programme and DFG (Bonn, Grant No. Le439/16).


\end{document}